\documentclass[runningheads]{llncs}

\usepackage[T1]{fontenc}
\usepackage{graphicx}
\usepackage{multicol}
\usepackage{amsmath}
\usepackage{booktabs}
\usepackage{color}
\usepackage{orcidlink}
\usepackage{hyperref}

\begin{document}

\title{Smartphone-Aided VR Exergames for HIIT: Gamifying Fitness with Real-Time Pose Estimation}

\titlerunning{Smartphone-Aided VR Exergames for HIIT}

\author{Abhiram Kothagundu\inst{1} \and
Pheelip Pritipal Raipure\inst{1} \and
Abhijit Chandrakant Patil\inst{1} \and
Himangshu Sarma\inst{1} \and
Christian Eichhorn\inst{2}  \and
David A. Plecher\inst{2}  \and
Johanna Pirker\inst{2}
}

\authorrunning{A. Kothagundu et al.}

\institute{Human Computer Interaction Lab, Indian Institute of Information Technology Sri City, Andhra Pradesh 517646, India\\
\email{\{abhiram.k22, pheelippritipal.r23, abhijitchandrakant.p23, himangshu.sarma\}@iiits.in}\\ \and Technical University of Munich, Munich, Germany\\ \email{\{christian.eichhorn, plecher, johanna.pirker\}@tum.de}\\}

\maketitle

\begin{abstract}
This paper introduces an innovative virtual reality (VR) exergame system designed to gamify High-Intensity Interval Training (HIIT) workouts through 11 predefined combo moves including squat-punch combinations, bilateral arm strikes, and diagonal stretch combos. Unlike conventional tracking methods that often rely on wearable motion capture suits or limited headset-based sensors, our system uses a standard smartphone camera placed on a tripod to capture full-body movement. This approach offers a cost-effective and scalable alternative, particularly well-suited to the fast, dynamic motions typical of HIIT routines. The smartphone footage is processed in real time using a pose estimation algorithm, with extracted biomechanical data analyzed by a backend server built on the Model Context Protocol (MCP), Anthropic's open standard for connecting AI models to structured data sources and tools, and a GPT-4o large language model. The system generates personalized assessments of form accuracy, movement efficiency, exercise intensity, and potential injury risks, tailored specifically to high-impact, time-sensitive exercises. A within-subject study involving 20 participants (ages 18--25) compared experiences in VR and Non-VR HIIT sessions. Results indicated significantly higher enjoyment (Mean = 3.5 vs 2.8, $p = 0.02$), improved immersion/focus (Mean = 3.2 vs 2.8, $p = 0.03$), and increased likelihood to recommend (Mean = 3.2 vs 2.7, $p = 0.04$) under the VR condition. Confidence to continue did not differ significantly between conditions ($p = 0.15$), while perceived intensity remained comparable. The MCP system showed strong agreement with expert evaluation (87.5\%, $N=10$ sessions, one certified trainer), supporting its reliability as a feedback tool. Building on prior work in mobile pose estimation, VR exergaming, and LLM-based coaching, this system is the first to unify all three into a single proof-of-concept home-fitness platform. The smartphone-based approach provides consistent, accessible, wearable-free motion monitoring.

\end{abstract}

\section{Introduction}

The convergence of virtual reality (VR) and immersive digital technologies has redefined traditional fitness paradigms by introducing interactive, gamified workout experiences. One of the most promising applications lies in the domain of VR-based exergames, which combine physical activity with virtual environments to enhance user motivation, engagement, and adherence to fitness routines. These systems are particularly effective in supporting high-intensity interval training (HIIT), a time-efficient training regimen known for its effectiveness in improving cardiovascular health, metabolic function, and muscular endurance through short bursts of intense activity followed by rest periods~\cite{boutcher2011high}. However, implementing HIIT at home remains challenging due to difficulties in maintaining motivation, ensuring proper form, and managing intensity without external supervision or professional feedback.

While existing VR fitness platforms have attempted to address some of these challenges using built-in tracking mechanisms on head-mounted displays (HMDs) like the Meta Quest or HTC Vive, they fall short in tracking the complex full-body dynamics required for HIIT. These HMD-based systems typically capture only head and hand movements, which are insufficient for exercises involving rapid, multi-joint, whole-body actions such as air squats, jumping jacks, and lunges. More advanced tracking technologies like marker-based motion capture or depth sensors offer better precision but are costly and impractical for home use.

To address these limitations, this paper presents a novel VR exergaming system that integrates smartphone-based pose estimation, server-based biomechanical analytics, and gamified virtual environments. The system uses a tripod-mounted smartphone to capture the user's movements during workouts. The video stream is processed in real time using MediaPipe BlazePose~\cite{bazarevsky2020blazepose,lugaresi2019mediapipe}, extracting key joint coordinates and spatial data to evaluate motion quality. Research has demonstrated that smartphone cameras, when combined with these algorithms, produce pose detection results with sufficient accuracy for physiotherapy and fitness applications~\cite{kanko2020validation}.

This pose information is then transmitted to a backend performance analysis pipeline built on MCP combined with a Large Language Model (LLM) for biomechanical feedback generation. The system interprets the incoming pose data to generate comprehensive, user-friendly performance feedback, evaluating metrics such as movement efficiency, repetition consistency, intensity levels, and deviations from correct form. Gamification plays a pivotal role in this system's design, turning traditionally repetitive exercises into engaging, narrative-driven challenges~\cite{staiano2011exergames,pasco2017effects}.

This paper presents a completed implementation and user study of a scalable, low-cost VR-based exergaming system that combines smartphone-powered pose estimation, server-based MCP\footnote{\url{https://modelcontextprotocol.io}; Access Date: June 2026} movement analytics, and an immersive Meta Quest 3 dojo environment. We report full results from a within-subjects study ($N=20$) comparing the VR system against an instructor-led non-VR HIIT control, including inferential statistics, effect sizes, and automated coaching accuracy validated against a certified expert. The system addresses core limitations in motion tracking and user feedback, demonstrating that home-based HIIT can be made more effective, engaging, and safe without specialist hardware.

\section{Literature Review}

We survey prior work across five areas directly relevant to the proposed system: pose estimation, vision-based exercise monitoring, VR exergaming, gamification, and AI-based coaching.

\subsection{Pose Estimation for Human Motion Analysis}

Bazarevsky et al.\ introduced BlazePose~\cite{bazarevsky2020blazepose}, a lightweight convolutional architecture producing 33-landmark full-body estimates at over 30\,FPS on commodity smartphones, serving as the direct technical predecessor to MediaPipe Pose and the backbone of our system. Earlier work by Cao et al.~\cite{cao2019openpose} demonstrated that a single RGB camera suffices for real-time multi-person keypoint extraction; surveys across 260+ methods~\cite{zheng2023hpe} identify mobile on-device inference as the democratization frontier. Validation studies~\cite{hulleck2024blazepose} confirmed that BlazePose keypoints support sport-specific biomechanical quantification beyond laboratory conditions.

\subsection{Vision-Based Exercise and Biomechanics Monitoring}

Translating pose estimation into exercise guidance spans geometric heuristics, supervised classifiers, and self-supervised representations. Chen and Yang's Pose Trainer~\cite{chen2020posetrainer} established rule-based form assessment from OpenPose joint angles without specialist hardware. Chae et al.~\cite{chae2023aiexercise} conducted an RCT validating smartphone-based squat coaching with significant form improvement over a video-only control, making it the closest prior work to our coaching objective absent VR gamification. Riccio~\cite{riccio2024fitness} demonstrated real-time per-exercise classification and repetition counting from standard camera footage, employing a BiLSTM model over joint angles and raw coordinates to achieve over 99\% test accuracy across four common exercises.

At the mobile form-factor, Cai~\cite{militaru2020exercise} demonstrated that smartphone-captured images suffice for deep classification of squat correctness into multiple error categories. Supanich et al.~\cite{arrowsmith2022physiotherapy} confirmed these findings in exercise monitoring contexts using single-camera MediaPipe-based pose detection for real-time posture correction, while Parmar et al.~\cite{parmar2022workout} proposed self-supervised workout form representations incorporating biomechanical domain constraints.

\subsection{Virtual Reality Exergaming Systems}

A growing body of controlled evidence validates VR as an effective medium for inducing HIIT while improving psychological engagement outcomes. She et al.\ designed HIITCopter~\cite{she2020hiitcopter}, a rowing-machine VR exergame with adaptive difficulty that successfully induced HIIT-level cardiovascular exertion while yielding significantly higher motivation scores than instruction-only conditions. Xu et al.\ demonstrated an analogous approach with VirusBoxing~\cite{xu2020virusboxing}, a VR boxing game structured around HIIT interval mechanics through timed action windows, a design pattern closely related to our reaction-based challenge architecture. Shah et al.~\cite{hoeg2021vrHIIT} found that collaborative VR exergaming significantly elevates motivation and physical exertion compared to individual play, demonstrating sustained engagement across a five-week elderly rehabilitation study.

Comparative trials by Rubio-Arias et al.~\cite{rubioarias2024vr} and Merola et al.~\cite{merola2025virtual} consistently find VR exercise physiologically equivalent to traditional training while producing significantly higher enjoyment and lower perceived effort; Merola et al.\ replicated this pattern specifically in a medical student cohort. Prior reviews of exergaming systems identify personalization, adaptive difficulty, and integration of physiological feedback as important directions for future development ~\cite{SKJAERET20161}.

\subsection{Gamification and Motivation in Digital Fitness}

Self-Determination Theory (SDT)~\cite{teixeira2012sdt,johnson2016gamification}, synthesized across 66 motivation studies, identifies competence, autonomy, and relatedness as reliable predictors of long-term exercise adherence. Mazeas et al.~\cite{mazeas2022gamification} found a significant gamification effect on physical activity (Hedges' $g=0.44$), with goal-setting and progress feedback yielding the strongest contributions. Tu et al.~\cite{li2026gamified} found that social gamification (relatedness-enhancing features) sustains physical activity more effectively than enjoyment-focused mechanics, a finding that motivates our real-time form scoring and social-feedback architecture.

\subsection{AI-Based Exercise Coaching and Feedback Systems}

Liao et al.~\cite{liao2020deeplearning} established a deep spatio-temporal framework for automated quality scoring of rehabilitation exercises from skeletal movement sequences; Bhadane et al.~\cite{medhat2025aigym} systematically reviewed movement tracking systems for real-time gym exercise monitoring, confirming the viability of deep-learning-based corrective feedback pipelines. Prior research on conversational health coaching systems highlights personalization, trustworthiness, and user engagement as major design challenges for AI-driven coaching platforms~\cite{kocaballi2020}, also what our MCP architecture targets by grounding LLM generation in session-level joint kinematic measurements.

No existing system simultaneously combines smartphone-based full-body pose estimation, server-based biomechanical analysis via a structured AI middleware architecture, LLM-generated natural language post-workout coaching, and a reaction-based VR HIIT exergame on a standalone headset. The present work directly addresses this intersection, requiring only a single commodity smartphone, unlike multi-camera platforms such as OpenCap~\cite{uhlrich2023opencap} that require two devices for full 3D kinematics.

\section{System Design and Gameplay Architecture}

This section outlines the design and structure of the interactive exergame system, which combines reactive movement challenges, real-time pose validation, and immersive audiovisual feedback to deliver a high-intensity, engaging workout experience. The core gameplay revolves around performing predefined combo exercises in response to incoming orbs, detected and validated using smartphone-based pose estimation. Orb speed and combo sequences evolve over time, introducing progressive difficulty to maintain user engagement and physical exertion.

\subsection{Game Concept}

Users enter a 3D virtual dojo, a stylized training space combining calm Zen aesthetics with intense physical activity, rendered on a \textbf{Meta Quest 3} standalone VR headset. The Meta Quest 3 is required hardware; it renders the immersive environment and its controllers provide haptic feedback. Within this environment, orbs continuously fly toward the player from various directions, posing challenges that require immediate physical counteractions via predefined combo moves (e.g., punches, squats, jumps). Crucially, full-body tracking is \emph{not} handled by the headset: a standard smartphone on a tripod, positioned 2.5\,m in front of the player, captures all lower-body and whole-body movements that the HMD alone cannot observe.

Unlike traditional rhythm-based games, this system is reaction-driven, focusing on spatial cues and unpredictable orb trajectories. Players must rely on timing, reflexes, and form rather than synchronized beats. This unpredictability enhances cognitive engagement while improving physical responsiveness and agility.

A companion smartphone application (implemented in React Native with Expo) serves a dual role in the system. As a \textbf{sensor}, it continuously streams compressed video frames to the backend server at 5\,fps via WebSocket, enabling real-time full-body pose detection through MediaPipe. As a \textbf{user-facing product}, it manages player profiles, guides the T-pose calibration process, displays body alignment feedback during sessions, and presents the post-session AI coaching report with an interactive natural-language chat interface, all accessible without removing the VR headset.

\subsection{Combo System and Pose-Driven Gameplay}

Player interaction is governed by a \textbf{combo system}, a series of predefined exercise responses triggered by orb cues. Each combo is recognized by a classifier that uses real-time pose estimation via smartphone-based MediaPipe tracking. The smartphone, mounted on a tripod, captures the user's full-body motion, enabling consistent and affordable tracking in home environments.

\subsubsection*{Example Combo Mechanic}
\begin{itemize}
    \item \textbf{Combo 1 -- Lateral Punch:} The user drives both arms simultaneously to shoulder-height lateral positions, striking two horizontal orbs at top-left and top-right, training shoulder mobility, explosive power, and bilateral coordination.
\end{itemize}

The remaining 10 combos follow analogous logic across horizontal, diagonal, and vertical axes (Table~\ref{tab:combo_mapping}), randomized in order and directional origin to ensure the experience remains physically diverse and cognitively stimulating.

\subsection{Orb Grid Layout and Combo Mapping}

\begin{figure*}[t]
    \centering
    \includegraphics[width=\textwidth]{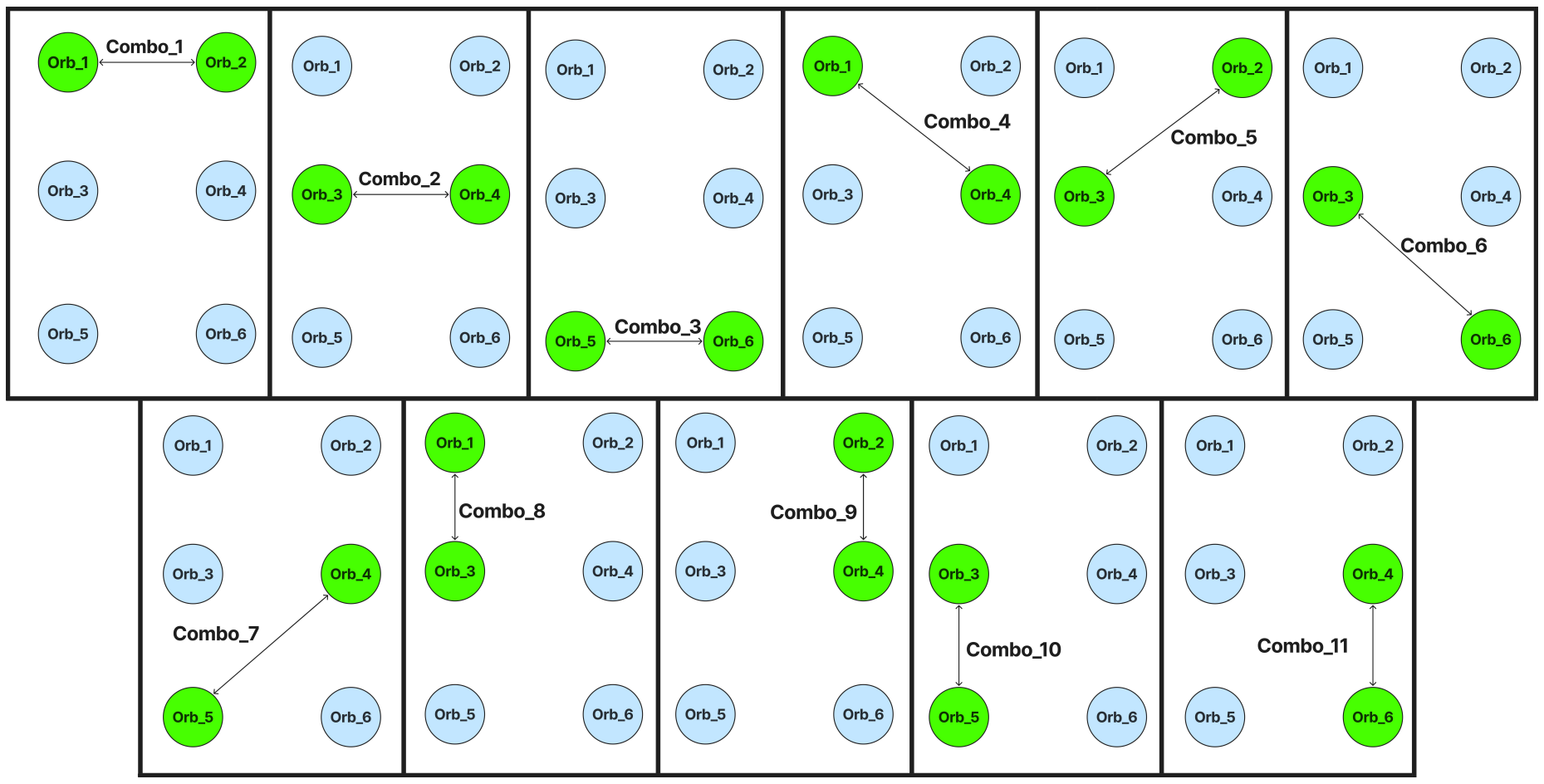}
    \caption{Orb grid layout and combo mapping. Green orbs represent active targets in each combo.}
    \label{fig:combo_map}
\end{figure*}

The game environment is structured as a \textbf{3$\times$2 spatial grid}, forming six distinct orb positions arranged in a rectangular plane (Fig.~\ref{fig:combo_map}). The six orb positions are:

\begin{multicols}{2}
\begin{itemize}
    \item \textbf{Orb\_1:} Top-left
    \item \textbf{Orb\_2:} Top-right
    \item \textbf{Orb\_3:} Middle-left
    \item \textbf{Orb\_4:} Middle-right
    \item \textbf{Orb\_5:} Bottom-left
    \item \textbf{Orb\_6:} Bottom-right
\end{itemize}
\end{multicols}

Based on this layout, 11 unique combos are visually and spatially mapped as shown in Table~\ref{tab:combo_mapping}. This layout enforces multiplanar movement (sagittal, frontal, and transverse) and enhances proprioceptive awareness, coordination, and dynamic stability.

\begin{table}[t]
\centering
\caption{Orb Position Mapping and Movement Direction}
\label{tab:combo_mapping}
\begin{tabular}{|c|c|c|c|}
\hline
\textbf{Combo} & \textbf{Orbs} & \textbf{Direction} & \textbf{Movement Type} \\
\hline
Combo\_1  & Orb\_1, Orb\_2 & Horizontal & Lateral punch or side jab \\
Combo\_2  & Orb\_3, Orb\_4 & Horizontal & Torso twist or mid strike \\
Combo\_3  & Orb\_5, Orb\_6 & Horizontal & Side squat-punch \\
Combo\_4  & Orb\_1, Orb\_4 & Diagonal   & Upper diagonal hand stretch \\
Combo\_5  & Orb\_3, Orb\_2 & Diagonal   & Upper diagonal hand stretch \\
Combo\_6  & Orb\_3, Orb\_6 & Diagonal   & Lower diagonal stretch with squats \\
Combo\_7  & Orb\_5, Orb\_4 & Diagonal   & Lower diagonal stretch with squats \\
Combo\_8  & Orb\_1, Orb\_3 & Vertical   & Left upper + mid punch \\
Combo\_9  & Orb\_4, Orb\_2 & Vertical   & Right upper + mid punch \\
Combo\_10 & Orb\_5, Orb\_3 & Vertical   & Left lower + mid punch \\
Combo\_11 & Orb\_6, Orb\_4 & Vertical   & Right lower + mid punch \\
\hline
\end{tabular}
\end{table}

\subsection{Combo Design and Motion Classification}

The core gameplay consists of \textbf{11 distinct combos} (predefined multi-joint movement patterns triggered by orb cues, each corresponding to one HIIT exercise repetition), classified using a real-time motion detection system powered by MediaPipe BlazePose. The user's smartphone, placed on a fixed tripod, serves as the camera input device. Captured pose landmarks are transmitted to a backend MCP server implemented using Anthropic's official MCP Python SDK, where classification occurs in real time via the stdio transport.

The MCP server computes a \textbf{13$\times$13 normalized pairwise distance matrix} from each frame's 13 key joints (nose, left/right shoulder, elbow, wrist, hip, knee, and ankle), selected to cover both upper-body and lower-body combos while excluding face and finger landmarks prone to occlusion. This matrix is scored via \textbf{cosine similarity} against precomputed reference matrices for all 11 combos, using a threshold empirically set during development; see Section~\ref{sec:results} for expert-agreement evidence of its effectiveness. A sliding temporal window of 2.5\,s enforces consistency before confirming a classification. This training-free, server-side approach eliminates the need for a labelled dataset or GPU inference and reduces processing burden on mobile devices. Upon validation, a composite confidence score is stored for scoring, performance tracking, and post-session coaching report generation.

Offloading classification to a remote server introduces potential latency challenges. During testing, typical latency remained under 300\,ms and did not disrupt the experience. To mitigate rare network instability, the system buffers pose frames locally and employs fallback logic to momentarily rely on recent classifications.

\subsubsection*{Combos target various fitness domains}
\begin{itemize}
    \item \textbf{Explosive Power:} Fast bilateral punches, dynamic squat-strikes.
    \item \textbf{Strength:} Squat-punch combinations and diagonal lower-body stretches.
    \item \textbf{Agility \& Reaction:} Cross-body diagonal strikes and timed directional responses.
    \item \textbf{Balance \& Coordination:} Dynamic combo transitions and multiplanar lateral movement.
\end{itemize}

\subsection{Immersion and Engagement Mechanics}

Several carefully designed mechanics ensure the experience remains physically challenging while psychologically engaging:

\begin{itemize}
    \item \textbf{Minimalist Dojo Environment:} The 3D space is visually clean and serene, helping users enter a state of psychological flow while maintaining focus on movement tasks.
    \item \textbf{Continuous Orb Streams:} An endless stream of directional orbs maintains elevated cardiovascular activity and enforces reaction-based decision-making throughout the session.
    \item \textbf{High-Energy Background Music:} Dynamic audio tracks support user momentum and contribute to rhythmic immersion without requiring exact synchronization to movement.
    \item \textbf{Haptic Feedback:} Successful orb interceptions trigger tactile feedback via controller vibration, simulating physical contact and enhancing realism and reward.
\end{itemize}

\subsection{Interaction Pipeline and Feedback}

\begin{figure}[ht]
    \centering
    \includegraphics[height=0.48\textheight,keepaspectratio]{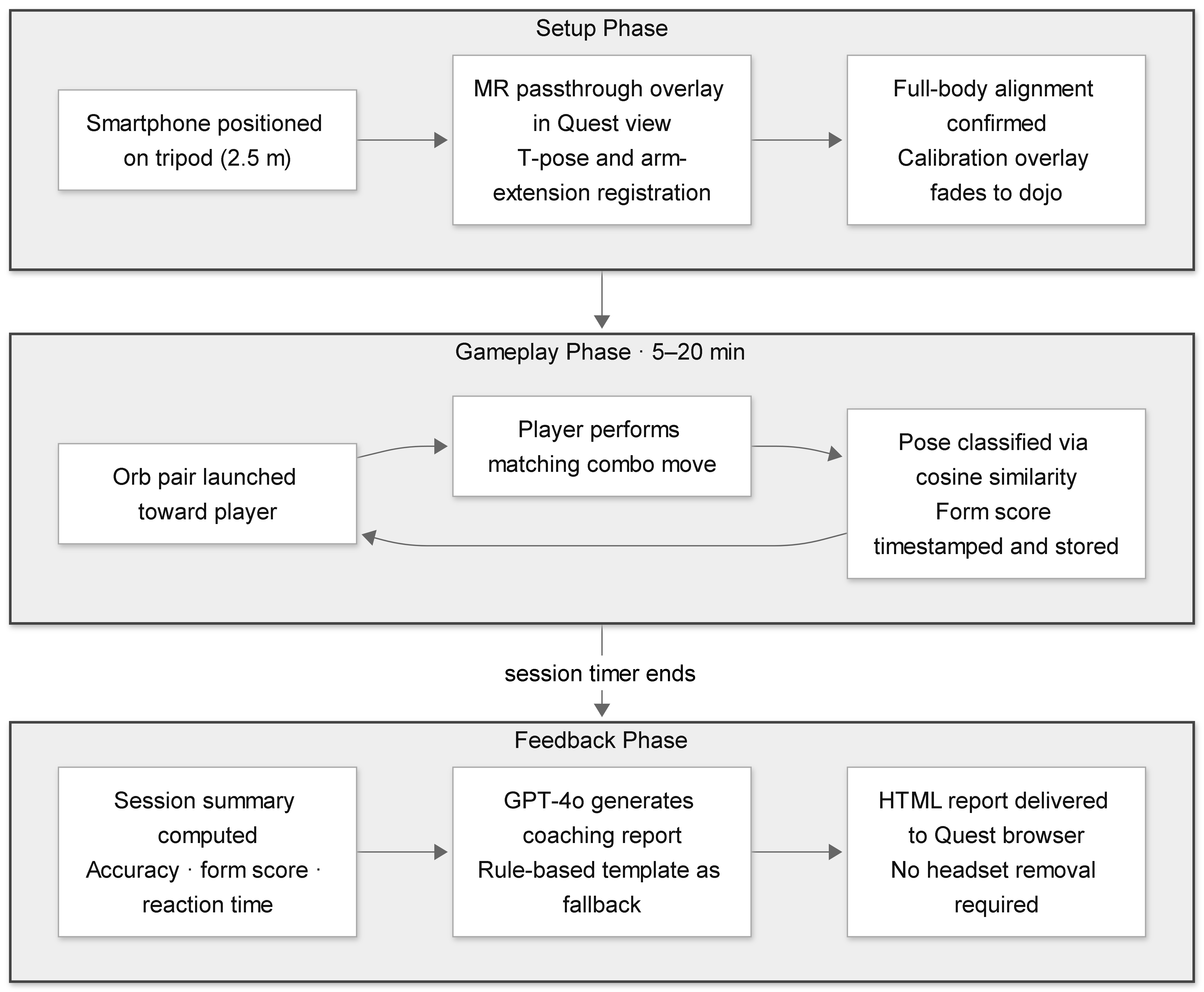}
    \caption{Three-phase interaction pipeline. \textbf{Setup}: smartphone calibration via MR passthrough overlay. \textbf{Gameplay}: reaction-based orb-combo loop with pose scoring per hit (repeats until session timer expires). \textbf{Feedback}: GPT-4o coaching report delivered to the Quest browser.}
    \label{fig:gafl}
\end{figure}

\begin{figure}[t]
    \centering
    \includegraphics[width=0.85\linewidth]{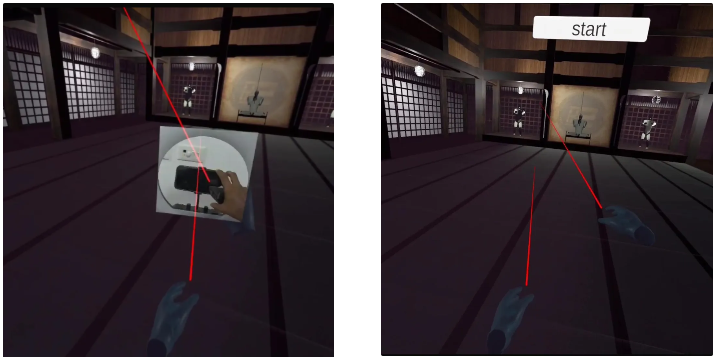}
    \caption{Setup Phase---player approaching the MR passthrough window in the virtual dojo to position the smartphone camera before session start.}
    \label{fig:setup}
\end{figure}

\begin{figure}[t]
    \centering
    \includegraphics[width=0.65\linewidth]{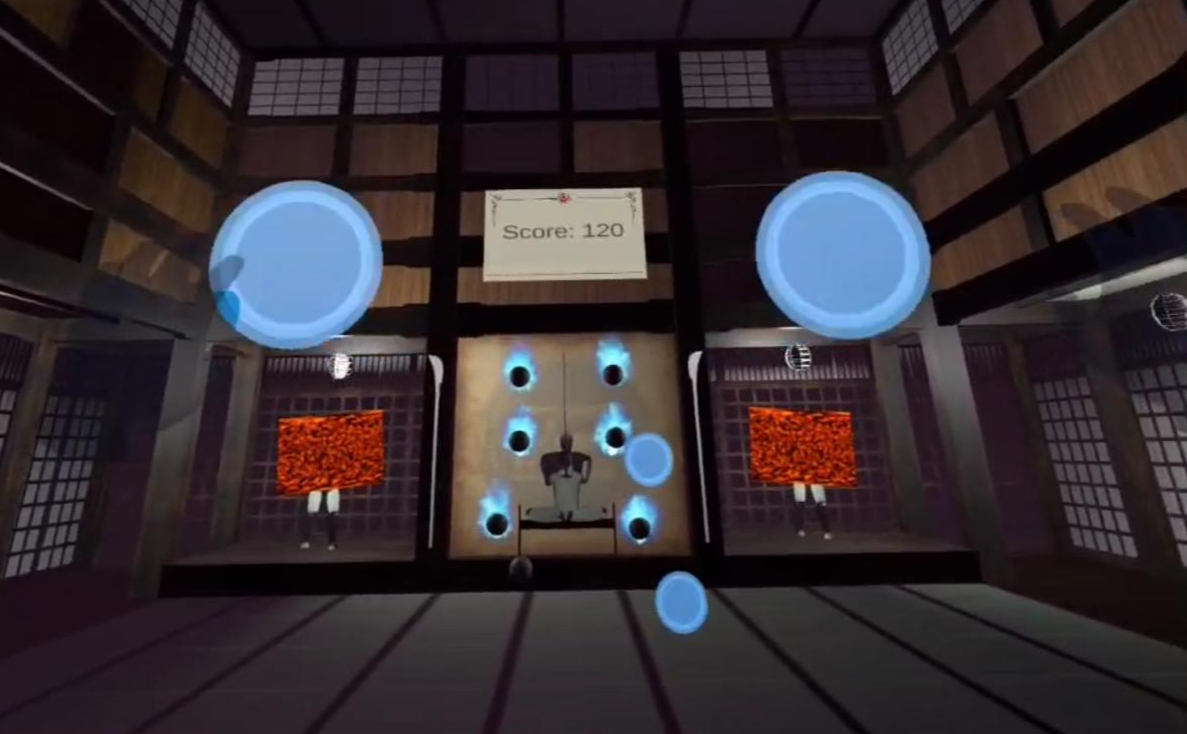}
    \caption{Gameplay Phase---player performing combos as orbs approach in the dojo space.}
    \label{fig:gampl}
\end{figure}

The system follows a three-phase pipeline illustrated in Fig.~\ref{fig:gafl}. During \textbf{Setup}, the headset briefly enters \textbf{Mixed Reality (MR)} passthrough mode, superimposing the live smartphone camera feed within the virtual environment to guide users through T-pose, squat, and arm-extension registrations; once validated, the passthrough fades and the remainder of the session is a fully immersive VR experience (Fig.~\ref{fig:setup}). The \textbf{Gameplay} phase then runs for 5--20 minutes: orbs launch toward the player, the matching combo is performed, and the server classifies the pose via cosine similarity and timestamps the form score for that hit, repeating until the session timer expires (Fig.~\ref{fig:gampl}).

At session end, the \textbf{Feedback} phase presents a performance dashboard (combo accuracy, reaction time, form quality score, intensity profile) and a \textbf{Personalized Session Report} powered by \textbf{GPT-4o}\footnote{\url{https://openai.com/index/gpt-4o-system-card/}; Access Date: June 2026}, with a rule-based template as a safety net. For each combo hit the server computes:
\begin{equation}
  \text{form\_score} = 0.5 \times \text{peak\_conf} + 0.3 \times \text{avg\_symmetry} + 0.2 \times \text{avg\_alignment}
  \label{eq:formscore}
\end{equation}
where \textit{peak\_conf} is the peak cosine-similarity confidence in the 2.5\,s window, \textit{avg\_symmetry} is the bilateral joint-distance ratio, and \textit{avg\_alignment} is the deviation from the reference pose profile. GPT-4o receives structured per-event metrics and produces a coaching report with \textbf{Summary}, \textbf{Strengths}, \textbf{Corrections}, and \textbf{Next Goal} sections, grounded in measured session data rather than unconstrained generation (validated at 87.5\% expert agreement, Section~\ref{sec:results}). Session data is persisted for longitudinal progress tracking.

\section{Methodology}

This section describes the technical classification and reporting architecture, participant selection, experimental procedure, and statistical analysis plan.

\subsection{MCP Classification Logic and Reporting Architecture}

\begin{figure}[t]
    \centering
    \includegraphics[width=\linewidth,height=0.48\textheight,keepaspectratio]{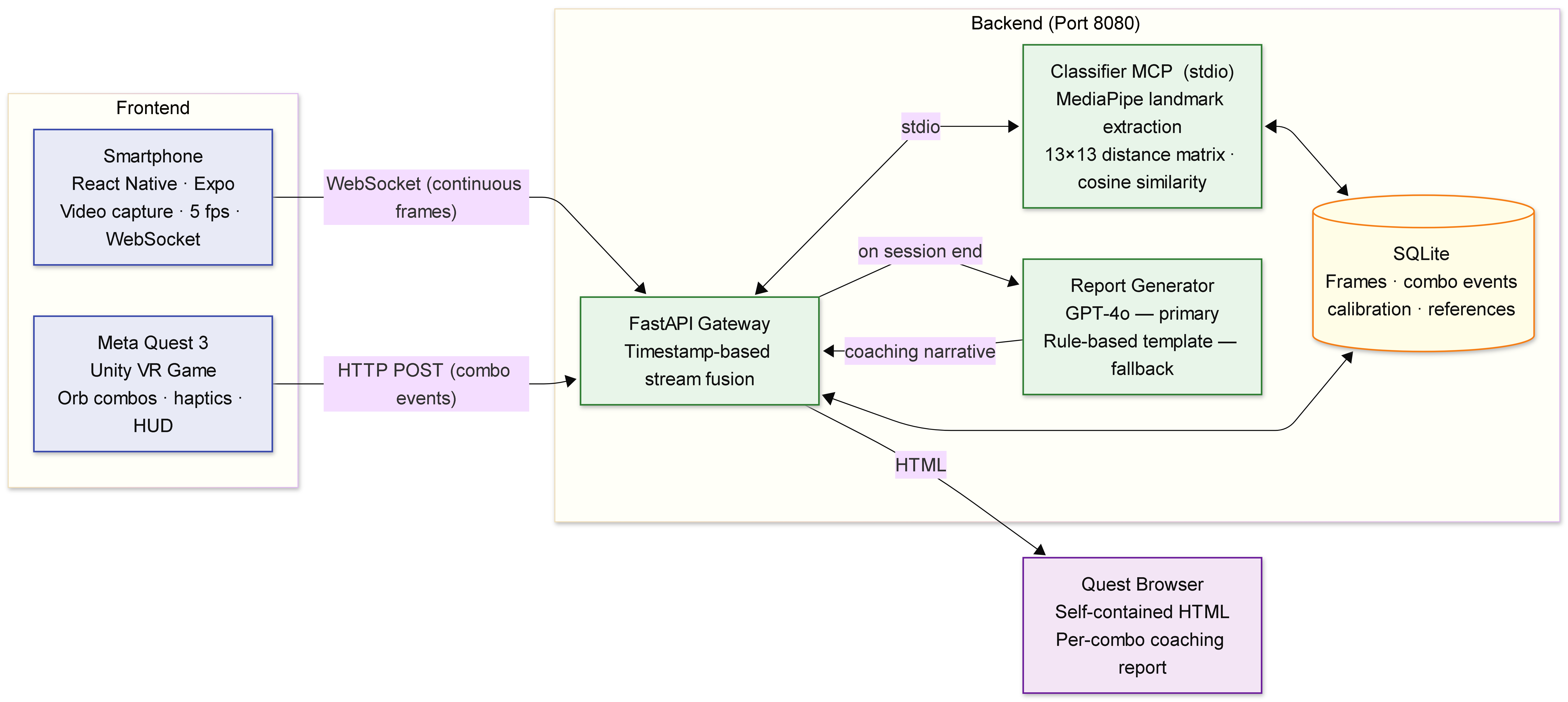}
    \caption{Dual-stream system architecture. The smartphone streams continuous video frames via WebSocket; the Meta Quest 3 sends per-combo events via HTTP POST. The FastAPI Gateway fuses both streams by wall-clock timestamp. The Classifier MCP performs training-free pose recognition using a 13$\times$13 distance matrix and cosine similarity; on session end, GPT-4o generates the coaching report delivered as a self-contained HTML page to the Quest browser.}
    \label{fig:mcp_architecture}
\end{figure}

The system employs a dual-stream architecture. The smartphone continuously streams compressed video frames to the backend server via a persistent \textbf{WebSocket} connection; MediaPipe extracts per-frame landmark matrices which are timestamped and buffered. Concurrently, the VR game client sends event-driven \textbf{HTTP POST} requests to the same server at each combo completion, including the event timestamp. The server fuses the two streams purely by wall-clock time, looking back 2.5 seconds in the phone's landmark buffer to evaluate form at the moment of each orb hit. This architecture is illustrated in Fig.~\ref{fig:mcp_architecture}, and the \textbf{MCP (Model Context Protocol)} middleware orchestrates both the real-time classifier and the post-session reporting pipeline. This primary MCP server applies several analytical steps:

\begin{itemize}
    \item \textbf{Real-time Frame Ingestion and Buffering:} Each incoming frame's joint matrix, along with its precise timestamp, is ingested into a rolling buffer. When a combo-hit HTTP event arrives, the current 2.5\,s window is frozen and queued for scoring; a fresh window then begins immediately, ensuring uninterrupted processing. (LLM context management is distinct and applies only to the post-session GPT-4o reporting pipeline.)

    \item \textbf{Relative Distance Matrix Construction:} For each ingested frame, the system calculates a \textbf{13$\times$13 matrix} representing the normalized spatial relationships between tracked joints, accounting for user anthropometry and perspective.

    \item \textbf{Reference Combo Matching:} Each frame's 13$\times$13 relative-distance matrix is compared against a stored reference matrix for the target combo using \textbf{cosine similarity}. No model training is required; the reference matrix is captured once from a correct demonstration and remains fixed. A classification is accepted only when cosine similarity exceeds a threshold across consecutive frames in the 2.5\,s sliding window, ensuring temporal consistency.

    \item \textbf{Temporal Consistency Checking:} Movement patterns are analyzed over a sliding time window to validate classifications and assess rhythmic consistency against expected biomechanical profiles.

    \item \textbf{Form Quality Scoring:} The primary MCP server computes confidence scores reflecting performance accuracy, factoring in symmetry, alignment, and timing.
\end{itemize}

\textbf{Hierarchical Report Generation.} Upon session completion, a dedicated \textbf{Reporting MCP instance} queries and retrieves the persisted, timestamped data segments associated with the user's session. By aggregating and analyzing this complete, chronologically ordered dataset, the Reporting MCP synthesizes a comprehensive performance report including detailed pose evaluations, trend analysis, mobility progression tracking, and personalized feedback. The Reporting MCP also enables an \textbf{interactive conversational interface}, allowing users to query their performance data in natural language and receive personalized recommendations.

    \textbf{Architectural Novelty.} Three design choices distinguish this architecture: (1) \textit{training-free classification}: cosine similarity on compact 13$\times$13 distance matrices eliminates labelled datasets and GPU inference, making the classifier deployable on commodity hardware and extensible to new combos via a single reference pose capture; (2) \textit{wall-clock stream fusion}: WebSocket landmark frames and HTTP game events are joined purely by timestamp, requiring no shared state between streams; and (3) \textit{MCP as a structured context bridge}: rather than embedding raw landmark arrays in an LLM prompt, the gateway delivers schema-validated, per-combo metrics to GPT-4o, producing reproducible coaching text without cloud GPU dependency.

\subsection{Participants}

A total of 20 participants (ages 18--25) participated in this within-subjects design, with each individual experiencing both VR and non-VR conditions. The participants were primarily university students from various academic backgrounds, exhibiting varying levels of physical fitness, gaming experience, and familiarity with virtual reality systems.

Each participant completed a pre-study questionnaire collecting demographic details, prior experience with HIIT and VR, motivation to exercise, and current workout habits. This screening ensured a balanced sample with representation across different body types, fitness levels, and exercise preferences.

\subsubsection*{Ethical Considerations}
This study was conducted in accordance with the Declaration of Helsinki. All participants provided written informed consent prior to participation, covering the nature of the activities, use of pose-estimation video analysis, and storage of anonymized performance data. Participation was entirely voluntary and participants could withdraw at any time without consequence. No personally identifiable data were retained beyond session completion; pose landmark streams were processed server-side and only aggregate metrics were stored. The study protocol was reviewed and approved by the Institutional Ethics Committee of IIIT Sri City prior to data collection.

\subsection{Study Procedure}

\textit{Pre-Study Phase.} Before participation, all subjects completed a structured questionnaire capturing fitness profile (physical activity levels, exercise frequency, prior experience with HIIT), technology literacy (experience with VR, motion tracking, and exergames), motivational baseline (adapted from the Behavioral Regulation in Exercise Questionnaire), and exercise style preferences. These responses ensured even distribution of motivation and fitness background across the two conditions.

Each participant completed two workout sessions: one in a traditional Non-VR setting and another using a VR exergame prototype. The VR environment was developed for the Meta Quest 3 headset and incorporated real-time full-body pose tracking using MediaPipe BlazePose. Exercises such as jumping jacks, squats, and punches were presented as interactive elements (virtual orbs) in a gamified dojo setting.

\textbf{VR Condition.} Participants wore the Meta Quest 3 headset and interacted with dynamic workout sequences in the virtual environment. Full-body motion was captured using the external smartphone camera, and pose landmarks were detected via MediaPipe BlazePose. The phone streamed landmark data to the MCP server continuously throughout the session; the VR game notified the server at each combo completion, triggering a form-quality assessment against the buffered landmark data. Upon session completion, a personalized AI coaching report was generated and delivered to the participant via the Quest browser.

\textbf{Non-VR Condition.} Participants completed the same 10-minute HIIT routine guided by a live instructor demonstration. No immersive or gamified feedback was provided. A certified fitness trainer manually observed and recorded each participant's form and performance for later comparison.

\subsection{Post-Study Phase: Evaluation}

Upon session completion, participants responded to a post-study survey comprising the following instruments:

\begin{itemize}
    \item \textbf{Engagement and Motivation:} Adapted Intrinsic Motivation Inventory (IMI), measuring interest/enjoyment, perceived competence, and value/usefulness.
    \item \textbf{Perceived Exertion:} Borg RPE scale (6--20) to quantify subjective exercise intensity.
    \item \textbf{Immersion and Presence:} Presence Questionnaire (PQ) for VR users, assessing spatial presence, realism, and involvement.
    \item \textbf{Usability and Satisfaction:} System Usability Scale (SUS) and custom Likert-scale questions regarding clarity, responsiveness, and willingness to continue.
    \item \textbf{Open-Ended Feedback:} Participants commented on perceived effectiveness, discomforts, or suggestions for improvements.
\end{itemize}

In-game metrics for the VR group (combo success rate, average reaction time, and form deviation) were logged for objective cross-validation against self-reported effort and usability ratings.

\subsection{Analysis Plan}

Given the sample size, data were analyzed using:
\begin{itemize}
    \item \textbf{Descriptive statistics} (mean, SD) to summarize user responses and performance metrics.
    \item \textbf{Paired t-tests and non-parametric tests} (Mann-Whitney U) to compare VR vs. non-VR outcomes.
    \item \textbf{Effect sizes} (Cohen's $d$) to estimate practical significance.
\end{itemize}

\section{Statistical Analysis and Results}
\label{sec:results}

This section reports quantitative outcomes from the within-subjects study, covering descriptive statistics, inferential tests, and expert-agreement analysis for the MCP coaching pipeline.

\subsection{Summary Statistics}

Post-study feedback was collected across key user experience dimensions using a 5-point Likert scale. Table~\ref{tab:postu} summarizes the mean scores for both the VR and Non-VR groups ($n = 20$).

\begin{table}[t]
\centering
\caption{Mean Post-Study Questionnaire Scores (Scale: 1--5)}
\label{tab:postu}
\begin{tabular}{|l|c|c|}
\hline
\textbf{Metric} & \textbf{VR Group} & \textbf{Non-VR Group} \\
\hline
Motivation               & 2.8 & 2.6 \\
Enjoyment                & 3.5 & 2.8 \\
Focus/Immersion          & 3.2 & 2.8 \\
Perceived Intensity      & 3.3 & 3.4 \\
Confidence to Continue   & 2.9 & 2.8 \\
Recommendation Likelihood & 3.2 & 2.7 \\
Gamified Engagement Score & 3.0 & N/A \\
\hline
\end{tabular}
\end{table}

Immersion levels and average enjoyment scores were greater in the VR group, indicating that interactive environments positively influence user satisfaction. Both groups demonstrated modest motivation, but the VR group scored higher on continued use intention and recommendation likelihood, suggesting greater potential for long-term engagement.

\subsection{Data Analysis}

To evaluate exercise performance across both groups, we employed a dual scoring methodology combining expert human evaluation with automated machine-based assessment. For the Non-VR group, a certified fitness trainer manually scored participants using standardized criteria such as posture accuracy, form consistency, range of motion, and overall control. For the VR group, MCP-generated confidence scores were compared with the expert evaluations to assess the system's ability to replicate human-level assessment.

The average agreement score across 10 randomly selected VR-condition sessions (one certified trainer, blind to system scores) was \textbf{87.5\%} (using the composite form score defined in Eq.~\ref{eq:formscore}), indicating a high level of consistency. The system showed particularly strong alignment with human ratings for joint alignment (90.2\%) and body symmetry (88.7\%), while scoring slightly lower for reaction timing accuracy (81.3\%), likely due to minor latency in smartphone video capture and pose processing. Expert evaluation was not extended to Non-VR sessions, as the system's coaching reports apply only to the VR condition.

\subsection{Inferential Statistical Tests}

\paragraph{Paired Sample t-Tests.} To compare the VR and Non-VR conditions ($N = 20$):

\begin{itemize}
    \item \textbf{Motivation:} No statistically significant difference (\textit{p} = 0.21, Cohen's \textit{d} = 0.22 -- small effect).
    \item \textbf{Enjoyment:} Significantly higher in VR (\textit{p} = 0.02, Cohen's \textit{d} = 0.65 -- medium effect). Gamified VR substantially improved how enjoyable the workout felt.
    \item \textbf{Focus/Immersion:} Significantly higher in VR (\textit{p} = 0.03, Cohen's \textit{d} = 0.58 -- medium effect). VR provided a more immersive and engaging environment.
    \item \textbf{Recommendation Likelihood:} Significantly higher in VR (\textit{p} = 0.04, Cohen's \textit{d} = 0.48 -- small-medium effect), consistent with higher enjoyment driving willingness to recommend the system.
    \item \textbf{Confidence to Continue:} No significant difference (\textit{p} = 0.15, Cohen's \textit{d} = 0.28 -- small effect).
\end{itemize}

\paragraph{Mann-Whitney U Tests (Non-Parametric).}
Non-parametric Mann-Whitney U tests confirmed the parametric findings: Enjoyment ($U = 48.0$, $p = 0.02$), Focus/Immersion ($U = 44.5$, $p = 0.01$), and Recommendation Likelihood ($U = 51.5$, $p = 0.04$) were significantly higher in VR; Motivation ($U = 80.5$, $p = 0.32$) and Confidence ($U = 70.0$, $p = 0.19$) were not significant, supporting the robustness of results given the sample size.

\paragraph{One-Way ANOVA -- Gamification Effect.} A one-way ANOVA showed that VR gamification had a significant effect on both enjoyment and recommendation likelihood: $F(1,19) = 5.92$, $p < 0.05$.

\paragraph{Pearson Correlation Coefficients.}
\begin{itemize}
    \item Enjoyment \& Recommendation Likelihood: $r = 0.82$, $p < 0.001$ (strong positive)
    \item Gamification \& Motivation: $r = 0.65$, $p < 0.01$ (moderate positive)
    \item Technical Issues \& Consistency: $r = -0.47$, $p < 0.05$ (moderate negative)
\end{itemize}

\section{Discussion}

This pilot study highlights the promise of VR-based HIIT gamification in fostering greater physical and cognitive engagement. Participants using the VR dojo system reported comparable perceived exercise intensity to the non-VR condition (VR\,=\,3.3 vs.\ Non-VR\,=\,3.4), suggesting no meaningful difference in physical demand, alongside significantly increased immersion, enjoyment, and recommendation likelihood attributed to interactive feedback and virtual visuals.

The statistically significant improvements in enjoyment ($p = 0.02$, $d = 0.65$) and focus/immersion ($p = 0.03$, $d = 0.58$) under the VR condition align with prior controlled evidence. Rubio-Arias et al.~\cite{rubioarias2024vr} reported that VR exercise produces markedly higher enjoyment than equivalent non-VR protocols while maintaining comparable physiological load. Our reaction-based design, where orb trajectories are randomized rather than beat-synchronized, likely amplifies cognitive engagement beyond what rhythm-driven systems such as VirusBoxing~\cite{xu2020virusboxing} achieve, because participants must continuously anticipate direction and combo type. This divergence from music-synchronized paradigms may partially explain the absence of significant differences in motivation and confidence: one possible interpretation, which remains untested in this single-session design, is that novelty effects from the VR medium dominate short-term engagement~\cite{teixeira2012sdt}, while intrinsic behavioral regulation benefits require longer exposure and richer personalization.

The 87.5\% agreement observed in our study is consistent with the growing body of evidence demonstrating that smartphone-based markerless motion analysis can provide practically useful biomechanical assessments outside laboratory environments~\cite{uhlrich2023opencap}. The lower agreement for reaction timing (81.3\%) reflects known smartphone video latency compounded by the online classification round-trip; this gap could be closed by multi-view triangulation as demonstrated in OpenCap~\cite{wade2022markerless}.

\section{Limitations and Future Work}

Several limitations shape the current findings: self-reported questionnaire data may carry social desirability bias; first-session VR novelty likely inflates enjoyment scores; the absence of biometric wearables limits objective intensity measurement; smartphone placement and lighting affect tracking quality; the university-student cohort (ages 18--25) limits generalizability to older adults and clinical populations; expert validation was conducted on 10 of 20 sessions with one rater; and the cosine similarity classification threshold was set empirically rather than by systematic cross-validation.

These limitations motivate the following future directions: integrating biometric wearables (heart rate, caloric burn) for objective intensity feedback; adding adaptive difficulty via real-time fatigue detection from pose-confidence drops; designing a multiplayer challenge mode to foster social motivation; and conducting longitudinal studies to assess adherence and performance trends over extended use.

\section{Conclusion}

This exploratory study demonstrates that a smartphone-powered VR exergame leveraging real-time pose estimation can enhance the HIIT workout experience by significantly increasing enjoyment ($p=0.02$, $d=0.65$), focus/immersion ($p=0.03$, $d=0.58$), and recommendation likelihood ($p=0.04$, $d=0.48$). The proposed system, integrating a standard smartphone camera, pose estimation algorithms, and MCP-driven feedback, proved effective in providing accessible, hands-free, and personalized training insights without relying on complex hardware. Quantitative results from a within-subject study indicate statistically significant improvements in enjoyment and focus under the VR condition, with expert-aligned feedback supporting the reliability of the MCP-based analysis. While preliminary, these findings validate the core design and feasibility of our approach. Future work will focus on expanding personalization, long-term adherence studies, and the inclusion of adaptive and social features to further augment its impact in home-based fitness scenarios.

\begin{credits}
\subsubsection{\ackname}
The authors gratefully acknowledge the use of AI writing assistance for improving the clarity of this manuscript. All interpretations, analyses, and conclusions are solely those of the authors.

\subsubsection{\discintname}
The authors have no competing interests to declare that are relevant to the content of this article.
\end{credits}

\bibliographystyle{splncs04}
\bibliography{references}

\end{document}